# High-pressure behavior of the Fe-S system and composition of the Earth's inner core

Z G Bazhanova, V V Roizen, A R Oganov

**Abstract.** Using evolutionary crystal structure prediction algorithm USPEX, we identify the compositions and crystal structures of stable compounds in the Fe-S system at pressures in the range 100-400 GPa. We find that at pressures of the Earth's solid inner core (330-364 GPa) two compounds are stable – $Fe_2S$ and FeS. In equilibrium with iron, only $Fe_2S$ can exist in the inner core. Using the equation of state of $Fe_2S$, we find that in order to reproduce the density of the inner core by adding sulfur alone, 10.6-13.7 mol.% (6.4-8.4 wt.%) sulfur is needed. Analogous calculation for silicon (where the only stable compound at inner core pressures is FeSi) reproduces the density of the inner core with 9.0-11.8 mol.% (4.8-6.3 wt.%) silicon. In both cases, a virtually identical mean atomic mass $\bar{M}$ in the range 52.6-53.3 results for in the inner core, which is much higher than $\bar{M} = 49.3$ determined for the inner core from Birch's law. For oxygen (where the relevant stable oxide at conditions of the inner core is $Fe_2O$) we find the matching concentration in the range 13.2-17.2 mol.% (4.2-5.6 wt.%), which corresponds to $\bar{M}$ in the range 49.0-50.6. Combining our results and previous works, we find that inner core density and $\bar{M}$ can be explained by only four models (in atomic %): (a) 86%(Fe+Ni) + 14%C, (b) 84%(Fe+Ni) + 16%O, (c) 84%(Fe+Ni) + 7%S + 9%H, (d) 85%(Fe+Ni) + 6%Si + 9%H, and some of their linear combinations (primarly, models (c) and (d)).

## 1. Introduction

Sulfur is one of the likeliest light alloying elements in iron-rich cores of terrestrial planets. In the Earth, the inner core is solid and outer core is liquid, and according to seismic models, the density of both inner and especially outer core is several percent lower than the density of pure iron or an iron-nickel alloy at relevant pressures and temperatures [1]. From Birch's law [2], one can deduce that the mean atomic mass in the core is approximately 49.3 [3], compared to 55.85 for pure iron. To explain these differences, one has to allow for approximately 10-20 mol. % of lighter elements – the prime candidates being S, Si, O, C and H [3]. Earlier, we have studied the Fe-C system and concluded that the presence of carbon can explain the density of the inner core [4]; here we consider the Fe-S system.

Poirier suggested [3] that sulfur remains a good candidate since it easily dissolves in iron. Li et al. [5] indicated that the Fe-FeS system exhibits eutectic behavior to at least 25 GPa and supported the view that the Earth's inner core contains a significant amount of sulfur. Chen et al. [6] determined the unit cell parameters of $Fe_3S$ using synchrotron X-ray diffraction techniques and externally heated diamond–anvil cells at pressures up to 42.5 GPa and temperatures up to 900 K. Their data suggest that ≈ 2.1 at. % (1.2 wt. %) sulfur lowers the density of iron by 1%. They estimated 12.5–20.7 at. % (7.6–13.0 wt. %) S in the outer core and 2.2–6.2 at. % (1.3–3.7 wt. %) S in the inner core, if compared to pure iron the outer core density is lower by 6–10% and inner core by 1–3%.

Alfè et al. [7] have calculated the chemical potentials of S, O and Si in liquid and solid iron and, assuming that the solid inner core and liquid outer core are in thermodynamic equilibrium, found that O partitions into the liquid much more strongly than S and Si and that the presence of O in the core is essential to account for inner-outer core density jump. Alfè et al. [7] proposed that the inner core contains 8.5 ± 2.5 at. % S and/or Si and 0.2 ± 0.1 at.% O and the outer core 10 ± 2.5 at.% S and/or Si and 8 ± 2.5 at.% O. Badro et al. [8] have calculated seismic wave velocities and densities in the (Fe–Ni, C, O, Si, S) system and compared them with seismic properties of the Earth's core. They found that oxygen is required as a major light element in the core, whereas silicon, sulfur, and carbon are not required. They also found that the concentration of silicon in the outer core cannot be higher than 4.5 wt,%, and the concentration of sulfur must be below 2.4 wt.%. Hirose et al. [9] have reviewed recent experimental and theoretical studies on the high *P-T* crystal structures of iron-nickel, -silicon, -oxygen, -sulfur, -carbon, and -hydrogen compounds. They wrote that, indeed, the inner core might include ~6 wt.% S if sulfur were the sole light alloying component. Moreover, considering the density jump across the inner core boundary they pointed out that outer core may include ~ 6 wt. % silicon, ~ 3 wt. % oxygen and 1-2 wt. % sulfur. Using experimental data on phase equilibria in the Fe-S system at pressures up to 200 GPa, Saxena and Ericsson [10] concluded that the inner core has a maximum temperature of 4428 (± 500) K with sulfur content of ~15 wt.%, but to be consistent with seismic data, presence of yet another light element is necessary.

To conclude, the status of sulfur as a candidate light alloying element in the Earth's core is somewhat uncertain. However, the existing evidence for and against sulfur in the core involves many assumptions and extrapolations. Here, we consider the Fe-S system, searching for stable iron sulfides at pressures of the inner core, analyzing their crystal chemistry and determining, on the basis of the most accurate available data, how much sulfur is needed in order to reproduce the observed density of the Earth's inner core.

Pure iron, sulfur and FeS have been subject of numerous experimental and theoretical studies and some degree of consistency is beginning to emerge. Akahama et al. [11], using angle-dispersive X-ray diffraction showed that a base-centered orthorhombic (bco) structure of compressed sulfur is stable above 83 GPa. The experimental results of Luo et al. [12], based on energy-dispersive X-ray diffraction at pressures up to 212 GPa, detected a transition to a new phase at 162 GPa. Its X-ray reflections have been indexed with a simple rhombohedral β-Po structure. Degtyareva et al. [13] studied sulfur up to 160 GPa and found an incommensurately modulated body-centered monoclinic (bcm) structure to be stable at 300 K and pressures between 83 and 153 GPa, above which it transforms into the β-Po structure. Pseudopotential *ab initio* calculations [14] were performed for three high-pressure phases of sulfur (bco, β-Po and body-centered cubic (bcc)). These calculations did not reproduce the reported bco to β-Po phase transition, but showed a transition from β-Po to bcc at ~550 GPa. First-principles calculations for sulfur at 200–600 GPa and $T = 0$ K within the local density approximation predicted phase transitions in the sequence: β-Po → simple-cubic (sc) → bcc, with the simple cubic structure favored over a wide range of pressures from 280 to 540 GPa [15].

Experimental and theoretical studies of iron at high pressures are numerous and only recently reached a consensus. It is established, both experimentally and theoretically, that at 13 GPa and room temperature, the non-magnetic hcp-phase of iron is stable [16-18]. At pressures relevant to the Earth's core (136-330 GPa for the outer core, 330-364 GPa for the inner core), several new phases have been proposed. For example, a double hexagonal close-packed phase of iron [19] and an orthorhombically distorted hcp structure [20] were reported at high pressures and temperatures; other proposals included a body-centered tetragonal structure and the bcc structure [21, 22]. The proposal of the bcc structure at ultrahigh pressures and temperatures presents an interesting case: there are two bcc phases at atmospheric pressure, one ferromagnetic (at ambient conditions) and one paramagnetic (with disordered local magnetic moments) at high temperatures [23]. In principle, a third, nonmagnetic high-pressure bcc phase is not impossible, but Stixrude and Cohen [24] found that this structure is dynamically unstable (i.e. possesses soft modes, that is, phonons with imaginary frequencies) at high pressures and zero temperature. Nevertheless, Belonoshko et al. [21] and Vocadlo et al. [25] demonstrated that soft modes in this structure are anharmonically suppressed at temperatures of the Earth's core. Belonoshko et al. [21], using an embedded-atom potential, found this phase to become thermodynamically stable, while Vocadlo et al. [25] used more accurate DFT calculations and found this phase to be slightly less stable than hcp – but this small free energy difference can be overturned by the addition of impurities (Si, S), which may stabilize the bcc phase. Following these works, Dubrovinsky et al. [26] also claimed on the basis of their experiments that the (Fe, Ni) alloy adopts the bcc structure at high pressures and temperatures – but experiments of Kuwayama et al. [27] found the opposite. In an experimental *tour de force*, Tateno et al. [28] showed that at actual conditions of the Earth's inner core pure Fe has the hcp structure. Calculations of Côté and

Vočadlo [29] showed that the addition of Ni to iron only further stabilizes the hcp structure over bcc.

Iron sulfide (FeS) has numerous phase transitions at low pressures. Its ultrahigh-pressure behavior was in detail investigated using first-principles calculations up to the pressure of 400 GPa [30], and the transition sequence at 0 K was found to be: troilite (FeS I) → antiferromagnetic MnP-type phase (FeS II) → monoclinic phase (FeS III) → nonmagnetic MnP-type phase (FeS VI) → *Pmmn* phase (FeS VII). The *Pmmn* phase with the distorted NaCl-type structure is stable from 135 GPa up to at least 400 GPa. Experimental results Sata et al. [31] showed that a CsCl-type structure is stable at high temperatures and pressures above 180 GPa, hinting at a possible problem with density-functional description of FeS. We address this issue here.

## 2. Methodology

Our calculations are based on the evolutionary crystal structure prediction method USPEX [32-34] and density functional theory [35, 36] within the generalized gradient approximation (GGA) [37]. Such calculations for pure Fe at high pressure [32] have produced the known lowest-enthalpy structure, hcp-Fe, in agreement with available experimental evidence [28]. For pure S we predicted ground-state structures by USPEX: the β-Po structure in the pressure range 100-540 GPa, and bcc at >540 GPa. From Fig. 1, one can see a direct transition of S from β-Po to bcc, bypassing the simple cubic structure, as in [14] and questioning the existence of the simple cubic structure predicted in [15]. We note, however, that at pressures around 350-400 GPa this phase is competitive with the β-Po phase (Fig. 1). Fig. 2 shows comparison between theoretical (at 0 K) and experimental equations of state of pure sulfur [11-13], iron [38-41], and FeS [42]. One can see good agreement between theory and experiment.

First, we have performed variable-composition evolutionary searches for stable compounds/structures in the Fe-S system at pressures of 300 GPa and 400 GPa, followed by detailed fixed-composition evolutionary searches at 300 GPa and 400 GPa for the most promising compositions (FeS, $Fe_2S$, $FeS_2$, $Fe_3S$, $FeS_3$, $Fe_3S_2$, $Fe_5S_2$ and $Fe_9S_5$). These searches were performed for FeS with 8 and 16 atoms per cell, for $Fe_2S$ with 6, 9, 12 and 16 atoms per cell, for $FeS_2$ with 6, 9, 12, 15, 18 atoms per cell, for $Fe_3S$ with 8, 12 and 16 atoms per cell, for $FeS_3$ with 4 and 8 atoms per cell, for $Fe_3S_2$ with 10, 15 and 20 atoms per cell, for $Fe_5S_2$ and $Fe_9S_5$ with 14 atoms per cell. A typical USPEX simulation included 30-40 structures per generation for fixed-composition runs (for variable-composition runs, the initial population included 150 structures, and all subsequent generations had 60 structures), the lowest-enthalpy 60% of which used for producing the next generation of structures (70% of the offspring produced by heredity, 10% by atomic permutation or transmutation, and 20% by lattice mutation). All structures produced by USPEX were relaxed using the GGA functional [37] and projector-augmented wave method [43, 44] as implemented in the VASP code [45]. We used PAW potentials with an [Ar] core (radius 2.3 a.u.) and [Ne] core (radius 1.9 a.u.) for Fe and S atoms, respectively, and plane wave kinetic energy cut-off of 600 eV. Structure relaxations done within USPEX simulations employed homogeneous Γ-centered meshes with reciprocal-space resolution of $2\pi \times 0.03$ Å$^{-1}$ and Methfessel-Paxton electronic smearing [46] with σ=0.07 eV. Having identified several lowest-enthalpy structures using USPEX, we carefully re-relaxed them and recalculated their enthalpies using the tetrahedron method with Blöchl corrections [47]. To assess the possible magnitude of electron correlation effects on phase stability, we compared results of GGA and GGA+U calculations, the latter being based on Dudarev's formulation of the DFT+U method [48].

Thermodynamic stability of compounds in the Fe-S system was determined using the thermodynamic convex hull construction: to put plainly, a compound is stable when it has lower free energy than any isochemical mixture of any other compounds. After thermodynamically stable phases were identified, we computed their phonons, to check their dynamical stability. Phonons and thermodynamic properties of Fe-S compounds were using the finite-displacement approach as implemented in the PHONOPY code [49, 50]. To perform phonon calculations, all structures were fully relaxed with a cutoff of 600 eV and relaxation proceeded until all force components became by absolute value less than 0.01 meV/Å. We constructed supercells (typically 2x2x2, with dimensions more than 10 Å) and displaced atoms by 0.01 Å to get the forces, which were then used to construct the force constants matrix. Then, the dynamical matrix was constructed and diagonalized at a very dense reciprocal-space mesh.

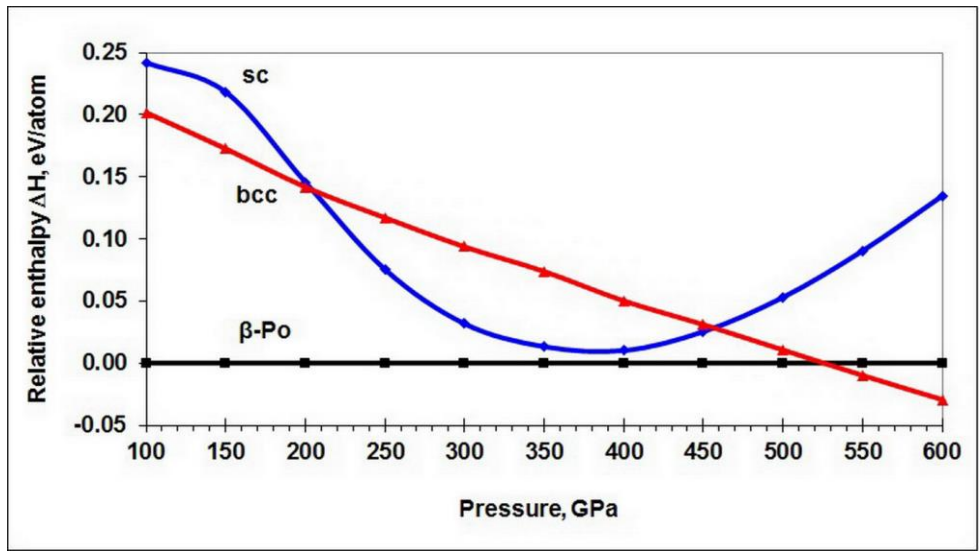

**Figure 1.** High-pressure stability of sulfur allotropes at T = 0 K.

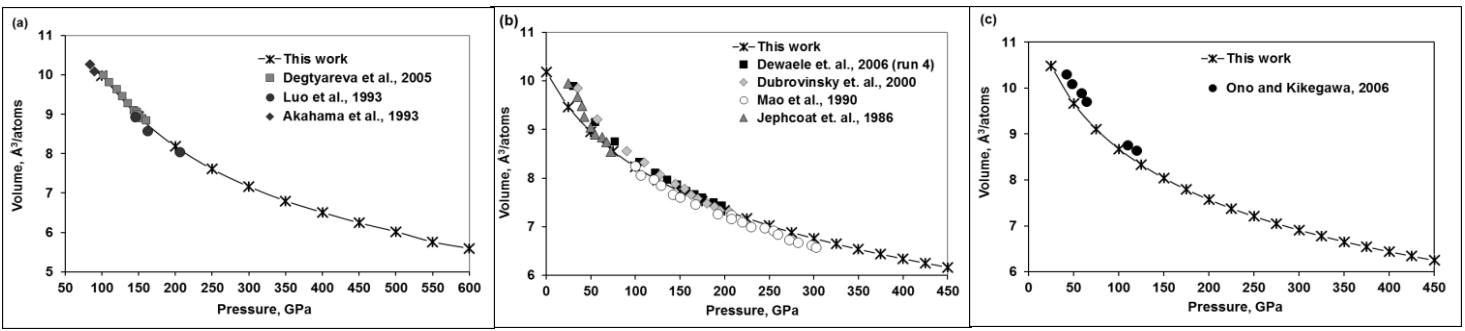

**Figure 2.** Comparison of the theoretical ($T = 0$ K) and experimental ($T = 300$ K) equations of state. (a) sulfur (experiments: Akahama et al. [11], Luo et al. [12], Degtyareva et al. [13]); (b) iron (experiments: Dewaele et al. [38], Dubrovinsky et al. [39], Mao et al. [40], Jephcoat et al. [41]); (c) FeS (experiments: Ono and Kikegawa [42]).

To clarify details of iron sulfide (FeS) high-pressure phase transitions we constructed a phase diagram in ($P,T$)-coordinates. For this purpose, we computed Gibbs free energies $G$ of the relevant FeS phases using quasiharmonic approximation:

$$G(P,T) = E_0(V) + F_{vib}(T,V) + P(T,V)V, \qquad (1)$$

where $E_0$ is the total energy from the DFT calculations and $F_{vib}$ is vibrational Helmholtz free energy calculated from the following relation:

$$F_{vib}(T,V) = k_B T \int_\Omega g(\omega(V)) \ln\left[1 - \exp\left(-\frac{\hbar\omega(V)}{k_B T}\right)\right] d\omega + \frac{1}{2}\int g(\omega(V))\hbar\omega d\omega, \qquad (2)$$

and pressure is

$$P(T,V) = -\frac{\partial(E_0(V) + F_{vib}(T,V))}{\partial V}. \qquad (3)$$

Here $g(\omega(V))$ is the phonon density of states at the given volume. Equations of state were fitted using the 3$^{rd}$-order Birch-Murnaghan equation of state [51]. By calculating differences of Gibbs free energies, we were able to construct the phase diagram.

## 3. Results

### 3.1. Structures and compositions of stable iron sulfides at ultrahigh pressures

Our predicted convex hulls at different pressures and the phase diagram of the Fe-S system are shown in Fig. 3. In the pressure range studied here, only three compounds have stability fields: $FeS_2$, FeS, and $Fe_2S$. Crystal structures of the new stable phases predicted by USPEX are given in Table 1. In the entire pressure ranges of their stability (Fig. 3b) the predicted phases are dynamically stable (Fig. 4).

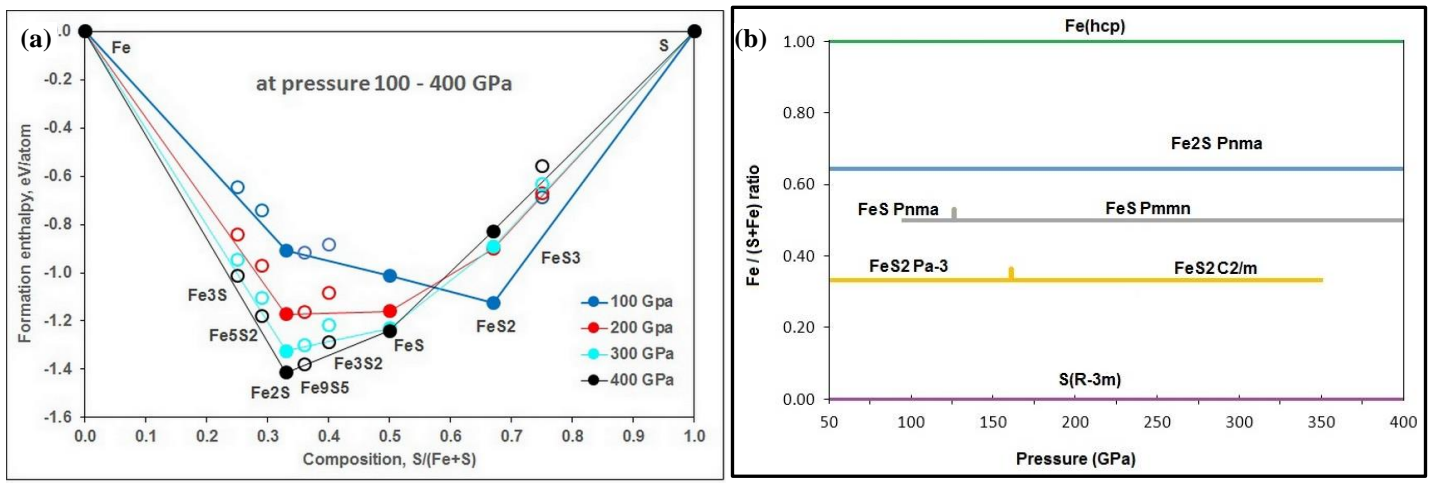

**Figure 3.** Thermodynamics of the Fe-S system: (a) convex hulls at different pressures and zero temperature and (b) composition-pressure phase diagram ($T = 0$ K).

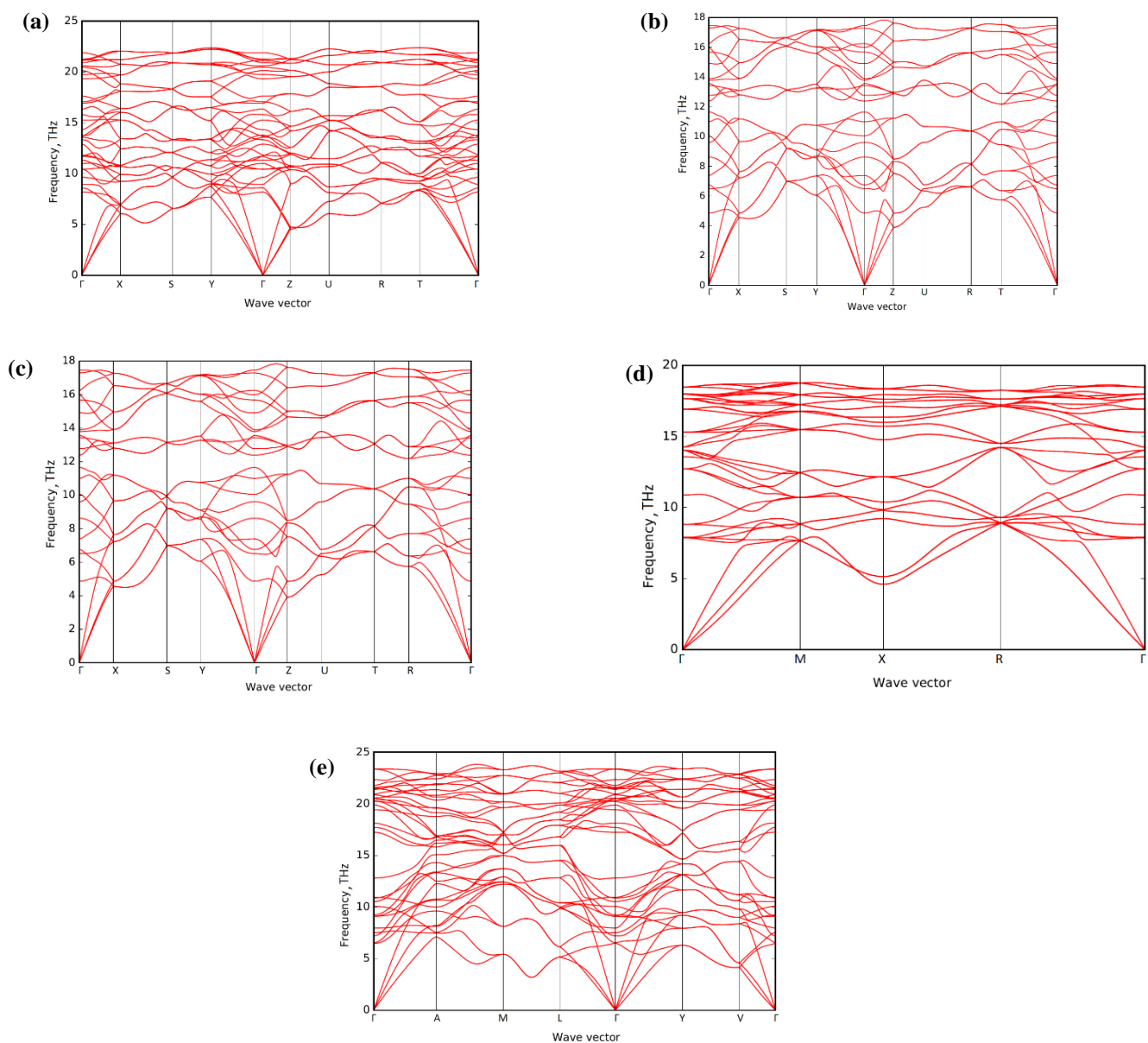

**Figure 4.** Phonon dispersion curves of (a) *Pnma*-Fe$_2$S at 200 GPa, (b) *Pnma*-FeS at 80 GPa, (c) *Pmmn*-FeS at 100 GPa, (d) *Pa*3-FeS$_2$ at 75 GPa, (e) *C*2/*m*-FeS$_2$ at 250 GPa. These structures are stable at 0 K.

In the whole pressure range studied here, Fe$_2$S has only one stable structure, which has *Pnma* space group and S atoms coordinated by ten Fe atoms in an irregular coordination. Geometrically, this is not a close-packed structure (Fig. 5a), but nevertheless its physical density is very high. Each Fe atom is coordinated by five S and eight Fe atoms.

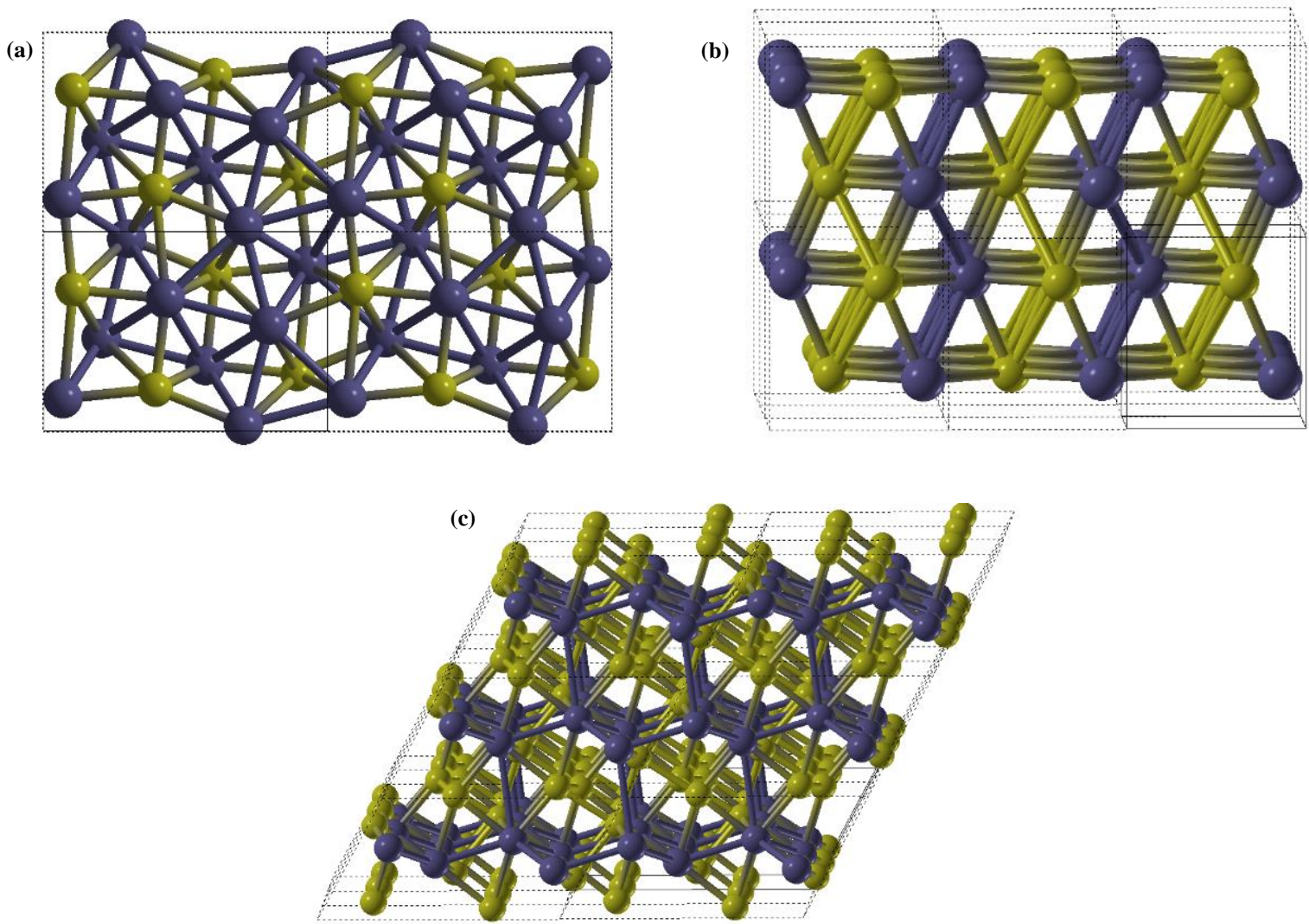

**Figure 5.** Structures of thermodynamically stable high-pressure iron sulfides found in this work. (a) – Fe$_2$S (*Pnma*), (b) - FeS (*Pmmn*), (c) – FeS$_2$ (*C2/m*). Purple balls – iron atoms, yellow balls – sulfur atoms.

**Table 1.** Structural parameters of some of the phases found by USPEX. *Pmmn*-FeS is the same phase as previously predicted by us (Ono et al. [30]) and confirmed here again.

| Phase (space group), pressure, unit cell parameters | Wyckoff position | | x | y | z |
|---|---|---|---|---|---|
| FeS (*Pmmn*), 300 GPa, a=3.629 Å, b=2.296 Å, c=3.317 Å | Fe | 2a | 0.75 | 0.75 | 0.8619 |
| | S | 2b | 0.25 | 0.75 | 0.6358 |
| Fe$_2$S (*Pnma*), 300 GPa, a=4.234 Å, b=3.178 Å, c=5.984 Å | Fe | 4c | 0.6665 | 0.25 | 0.4304 |
| | Fe | 4c | 0.4687 | 0.75 | 0.2053 |
| | S | 4c | 0.7885 | 0.75 | 0.6052 |
| FeS$_2$ (*C2/m*), 400 GPa, a=6.582 Å, b=3.624 Å, c=3.766 Å, β=118.41° | Fe | 4i | 0.6338 | 0.5 | 0.4027 |
| | S | 4i | 0.5331 | 0 | 0.2798 |
| | S | 4i | 0.7996 | 0 | 0.0332 |

In the pressure range investigated here (100-400 GPa), FeS has two stable phases: *Pnma*-phase (well-known MnP-type structure, a distortion of the NiAs structure type), stable below 122 GPa, and *Pmmn*-phase, which is stable above 122 GPa. The structure of the *Pmmn*-phase (Fig. 5b) can be considered as a strongly distorted NaCl-type structure, with well-defined close-packed layers of Fe and S atoms.

In the same pressure range, FeS$_2$ has two stable phases: well-known *Pa*3-phase (pyrite), predicted to be stable at pressures below 188 GPa, and *C2/m*-phase, stable above 188 GPa. Remarkable structure of *C2/m*-FeS$_2$ (Fig. 5c) has a diamond-like Fe sublattice (each Fe is surrounded by four other Fe atoms and six S atoms). Each S has three neighboring Fe atoms and three neighboring S atoms. At low temperatures and low pressures (below 6 GPa according to experiment of Parthasarathy [52], and below 3.7 GPa according to GGA calculations of Gudelli [53] another polymorph, marcasite (space group *Pnnm*), is stable.

In high-pressure high-temperature experiments, the following phases of FeS have been observed: *Pnma* at pressures at least up to 120 GPa [42], and CsCl-type (space group *Pm3m*) at pressures above 180 GPa [31]. Our calculations (Fig. 6) show the *Pnma* → *Pmmn* phase transition at 122 GPa (GGA) or 192 GPa (GGA+U, with U=4 eV), and both with and without the U-correction, the CsCl-type phase is always metastable at zero Kelvin temperature (by 0.1-0.15 eV/atom, which can be overturned by thermal effects, see below). The difference between GGA and GGA+U with such a large value of U gives an upper bound of the effects of electron correlation, because these effects are strongly reduced in metallic materials, such as FeS phases.

The CsCl-type phase, observed in experiment, is most likely stabilized by temperature. We have computed the high-pressure phase diagram of FeS based on GGA calculations and quasiharmonic approximation (Fig. 7). We see indeed that the CsCl-type phase may have a stability field at high temperatures. Experimental points where this phase was found (186 – 270 GPa, 298 – 1300 K) [31] are in the computed stability field of the *Pmmn*-phase, but even a moderate change of thermodynamic properties (to compensate for the errors of the GGA and of the quasi-harmonic approximation) of phases can extend the stability field of the CsCl-type phase to include these *P,T*-conditions.

The predicted [30] *Pmmn* structure of FeS is quite remarkable due to its unexpected relationship with one known polymorph of FeS – mackinawite. We found (Fig. 8) that upon decompression to 20 GPa our *Pmmn* structure spontaneously, i.e. without any activation barrier, transforms into a distorted mackinawite structure (space group *Pmmn*), which becomes ideal (undistorted) mackinawite (space group *P*4/*nmm*) at 15 GPa. This metastable transition is barrierless, first-order, and has a 11.3% density jump. The mechanism of the mackinawite-*Pmmn* transition involves the collapse and distortion of tetragonal mackinawite layers and formation of interlayer bonds. This structural relationship implies that *Pmmn*-FeS can be made relatively easily by compression of mackinawite.

At to 0 GPa mackinawite has theoretical cell parameters *a*=*b*=3.596 Å and *c*=5.794 Å; experimental values are *a*=*b*=3.674 Å and *c*=5.033 Å [54]: one can see good agreement for *a*=*b* parameters, while there is a significant discrepancy for the *c*-parameter, due to neglect of van der Waals interactions at the GGA level of theory. Indeed, the *c*-parameter corresponds to the interlayer distance, entirely determined by the van der Waals (vdW) interactions. We conducted relaxation of the mackinawite structure using two methods including vdW-correction:

D2 method of Grimme (DFT-D2 [55]) and Tkatchenko-Scheffler method (DFT-TS [56]). As well, we tested the vdW-DF correlation functional of Langreth and Lundqvist et al. (optB-PBE [57-61]) together with optPBE exchange-correlation functional which was optimized for use with vdW-DF functional. Experimental and theoretical lattice parameters of mackinawite are presented in the Table 2. One can notice that in particular optPBE-vdW method produces results in an excellent agreement with experimental data.

**Table 2.** Experimental [54] and theoretical (this work) lattice parameters of mackinawite at 0 GPa

|  | a | b | c |
|---|---|---|---|
| Experimental | 3.674 | 3.674 | 5.033 |
| GGA | 3.596 | 3.596 | 5.794 |
| GGA+D2 | 3.562 | 3.562 | 4.900 |
| GGA+TS | 3.570 | 3.570 | 5.074 |
| optPBE-vdW | 3.602 | 3.602 | 5.026 |

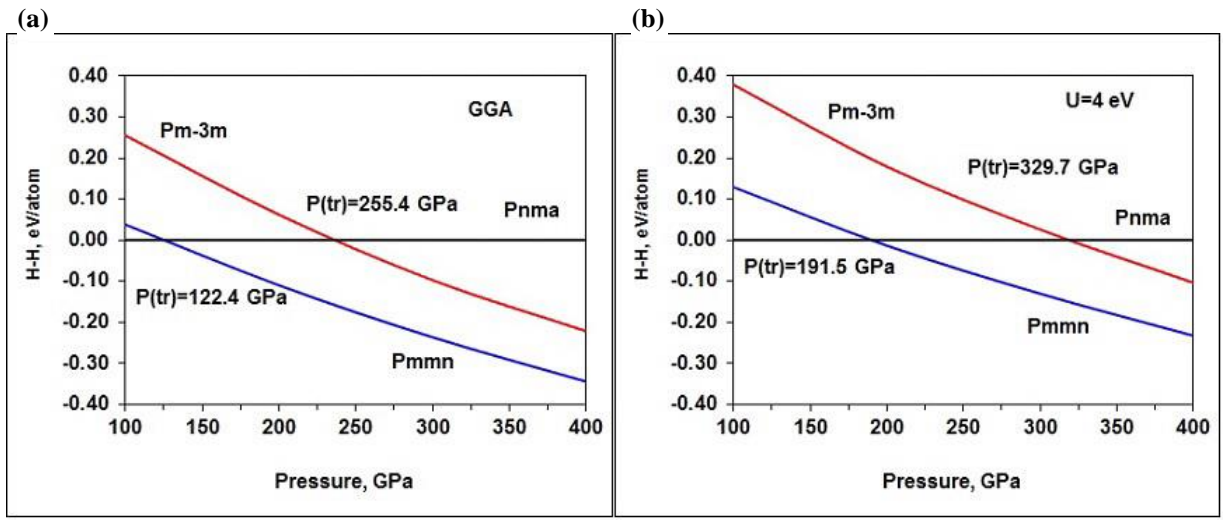

**Figure 6.** Enthalpy differences between FeS phases. (a) GGA calculations, (b) GGA+U (U=4 eV) results at $T = 0$ K.

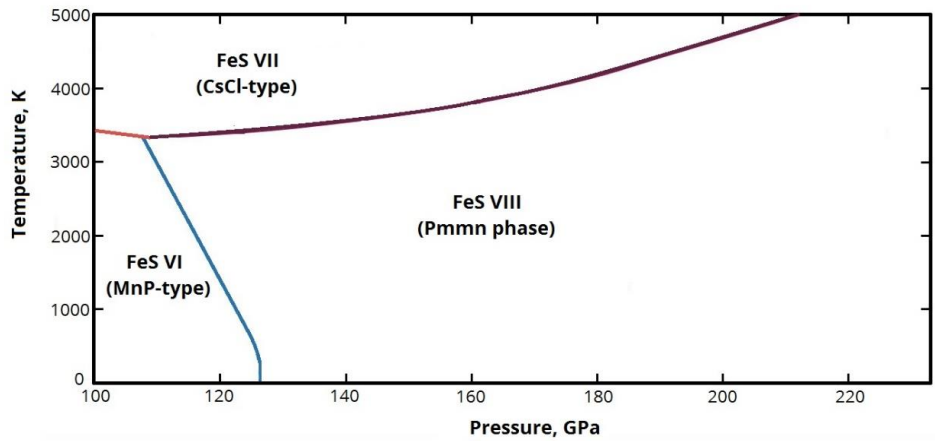

**Figure 7.** Phase diagram of FeS obtained in the quasiharmonic approximation.

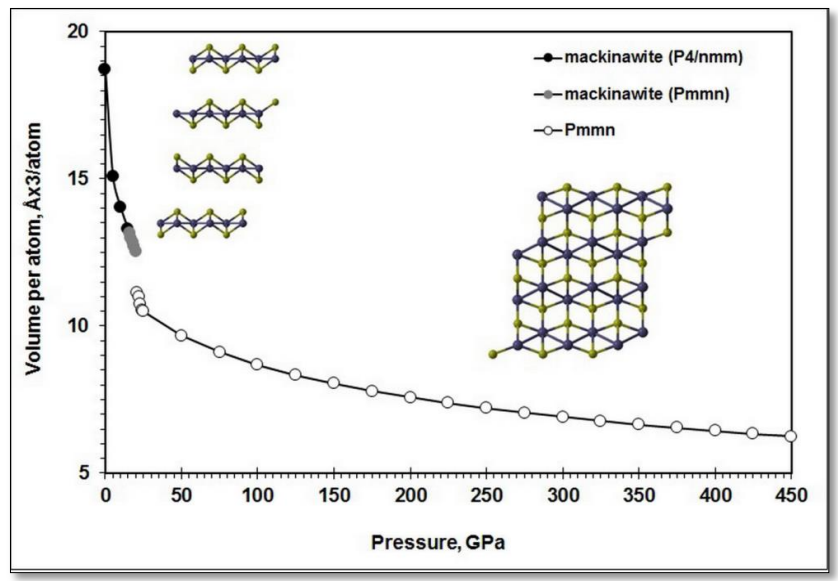

**Figure 8.** Decompression behavior of the predicted *Pmmn* phase of FeS at 0 K, showing a spontaneous transition to mackinawite, and a close relationship between the two structures (GGA results).

### 3.2. How much sulfur is needed to explain the density of the inner core?

Parameters of the predicted static equations of state of Fe-S phases are given in Table 3. Using the computed equations of state, we determined the molar concentration of sulfur needed to explain the density of the inner core by matching the observed density of the inner core to the density of the mixture of hcp-Fe and the stable sulfide $Fe_2S$. We also tried FeS - although this compound is unlikely at realistic concentrations of sulfur in the inner core, such a calculation is important for estimating robustness of our matching concentrations, i.e. their independence on the reference phase used. For comparison and for completeness, we also computed the matching concentration of silicon, using previous calculations for the Fe-Si [62] and Fe-O systems [63], see Table 4.

**Table 3**. Theoretical third-order Birch-Murnaghan equations of state of the non-magnetic high-pressure phases in the Fe-S systems.

| Phase | $V_O$, Å3/atom | $K_o$, GPa | $K_o'$ |
|---|---|---|---|
| *hcp*-Fe | 10.16 | 303.37 | 4.31 |
| *Pmmn*-FeS*) | 11.71 | 170.02 | 4.50 |
| *Pnma*-Fe2S | 10.69 | 210.11 | 4.74 |
| *C2/m*-FeS2 | 13.51 | 104.05 | 4.56 |
| *Pm3m*-FeSi | 10.46 | 256.0 | 4.25 |
| *R-3m*-S | 17.11 | 66.71 | 3.82 |

*[)] Parameters of the 3[rd]-order Birch-Murnaghan equation of state in the pressure interval 25–450 GPa. Results are very close to the published theoretical values (Ono et al. [30]): $V_0$=11.74 Å$^3$/atom, $K_0$=176.0 GPa, $K_0$'=4.35.

**Table 4.** Estimated matching concentrations of sulfur, silicon, carbon, hydrogen and oxygen based on reference stable binary compounds FeS, Fe$_2$S, FeSi, Fe$_2$C, FeH and Fe$_2$O at inner core conditions.

| Fe - X | mol.% X | wt.% X | $\overline{M}$ |
|---|---|---|---|
| *Pmmn*-FeS | 9.5-12.4 | 5.7-7.5 | 52.9-53.6 |
| *Pnma*-Fe$_2$S | 10.6-13.7 | 6.4-8.4 | 52.6-53.3 |
| *Pm-3m*-FeSi | 9.0-11.8 | 4.8-6.3 | 52.6-53.3 |
| *Pnma*-Fe$_2$C | 11.2-14.6 | 2.6-3.6 | 49.4-50.9 |
| *Fm-3m*-FeH | 16.9-22.0 | 0.4-0.5 | 43.8-46.6 |
| *I4/mmm*-Fe$_2$O | 13.2-17.2 | 4.2-5.6 | 49.0-50.6 |

These estimates give the concentration needed to match the inner core density if sulfur (or silicon) were the only light alloying element and, since several alloying elements are likely to the present in different concentrations, gives the upper bound for the concentration of each element. Such procedure of the calculation of the molar concentration of the light element for

Fe-C and Fe-H alloys was carried out in our earlier paper [4]. It is crucial to incorporate the effects of temperature, for which we use the formula:

$$\rho_{IC} = \rho_{Fe}^{T} + \frac{\partial \rho}{\partial x} \cdot x \Rightarrow \rho_{IC} - \rho_{Fe}^{T} = \frac{\rho_{Fe_2S}^{0K} - \rho_{Fe}^{0K}}{0.33} \cdot x, \qquad (4)$$

where $\rho_{IC}$ is the PREM (Preliminary Reference Earth Model) [64] density of the inner core at each depth, $\rho_{Fe}^{T}$ is the density of pure iron at given temperature [38], $\rho_{Fe}^{0K}$ and $\rho_{Fe2S}^{0K}$ are the computed zero-Kelvin densities of Fe and $Fe_2S$. The number 0.33 in Eq. (1) indicates the molar fraction of sulfur in $Fe_2S$. The resulting concentrations for the Fe-S and similarly made estimates for the Fe-Si and Fe-O systems (from this work) and Fe-C and Fe-H alloys (from Bazhanova et al. [4]), estimated along to isotherms, 5000 K and 6000 K, are given in Table 4.

Three conditions must be met if a certain element (S, Si, O, C and H) is to be considered as the main light elements of the core:

(i) the concentration of the light elements needed to explain the observed core density at the expected core temperatures (5000 ± 6000 K [65]) should not be unacceptably large (roughly, < 20 mol.%);

(ii) this amount should not display large and non-monotonic variations with depth, and

(iii) the resulting mean atomic mass $\bar{M}$ should be reasonably close to the one determined from Birch's law [2], i.e. 49.3 [3].

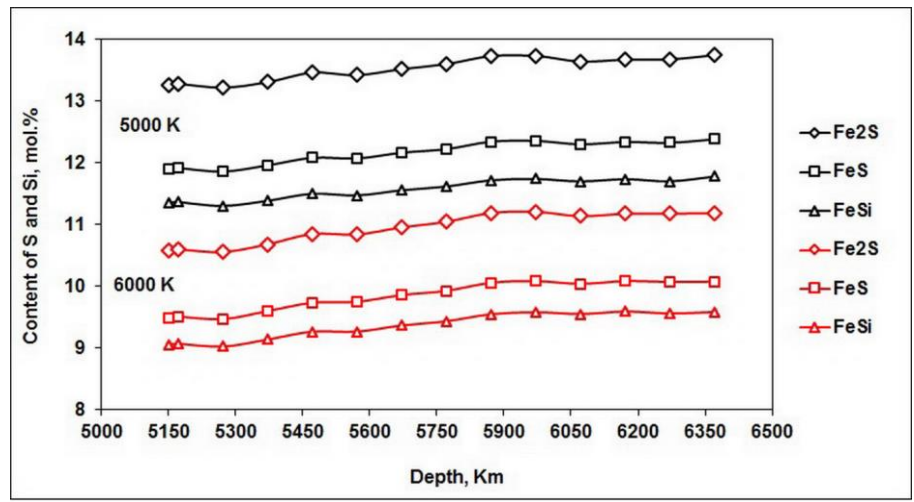

**Figure 9.** Matching molar concentrations of sulfur and silicon as a function of depth in the inner core using equations of state of $Fe_2S$, FeS and FeSi. Results are traced along the 5000 K and 6000 K isotherms.

For both Fe-S and Fe-Si systems (Fig. 9) we see mild variations of the matching concentrations with depth, leading to the average atomic mass $\bar{M}$ ~53, which is much higher than $\bar{M}$ =49.3 [3] derived for the Earth's core from Birch's law [2]. Neither silicon nor sulfur can be the only light alloying element. Likewise, hydrogen also cannot be the only light alloying element: its matching concentration shows more variation with depth [4], and corresponds to the average atomic mass $\bar{M}$ is ~45, much too low. Carbon and oxygen are the only elements that alone can explain both the density and $\bar{M}$ of the inner core with the composition (in atomic %):

86%(Fe+Ni) + 14%C, or

84%(Fe+Ni) + 16%O.

Oxygen is known to have very low solubility in crystalline iron at pressures and temperatures of the Earth's core [7] – but is highly soluble in liquid iron at conditions of the outer core [7] and can form in the inner core a separate phase (e.g., $Fe_2O$ [63]) precipitating from the molten outer core (note, however, that recent experiments cast doubt on the presence of large concentrations of oxygen even in the outer core [66]).

Considering simultaneous presence of two light alloying elements, we found that only two models produced satisfactory results:

84%(Fe+Ni) + 7%S + 9%H,

85%(Fe+Ni) + 6%Si + 9%H

It is remarkable that all these models have practically identical iron (+nickel) content of ~85%. We also note that the latter two models can be linearly mixed in any proportion, but not all other models can be combined: it was shown [67,68] that hydrogen and carbon cannot coexist in the Earth's core, i.e. are mutually exclusive. Silicon and oxygen are also mutually exclusive [69]. We do not know whether carbon and oxygen are mutually compatible in the core.

Checking against all the other known properties of the core (compressional and shear wave velocities, Poisson's ratio, density jump at the inner-outer core boundary) is necessary as it will help to discriminate between the four compositional models found here. It is already known that the carbon-only model, with ~14 mol.% carbon, explain not only the density and the Birch number of the inner core, but according to recent experiments [70-72] also the compressional and shear wave velocities and the anomalous Poisson ratio of the inner core. Indeed, Chen et al. [71] favored this model and concluded that up to two thirds of Earth's carbon can be in the Earth's core. Now the other models need to be tested against these constraints as well.

**4. Conclusions**

Using evolutionary crystal structure prediction, we have studied the Fe-S system at pressures from 100 GPa to 400 GPa, which encompasses the entire pressure range of the Earth's core. We found that only three compounds are stable – $Fe_2S$, FeS, and $FeS_2$ – and analyzed their crystal structures. We resolve the FeS conundrum, where the experimentally observed high-pressure high-temperature CsCl-type phase was not found to be stable in theory; here we show that this phase is thermally stabilized and does not have a stability field at zero Kelvin. We show that the predicted high-pressure *Pmmn*-phase of FeS is structurally related to well-known mackinawite and can be easily obtained by compression of mackinawite. Among the predicted compounds, in equilibrium with excess Fe, only $Fe_2S$ may exist in the Earth's inner core. We determine the amounts of S, Si and O, needed to explain the observed density of the inner core, and find them to be around 12, 10, and 16 mol.%, respectively. We arrived at four plausible compositional models of the inner core (in atomic %): (a) 86%(Fe+Ni) + 14%C, (b) 84%(Fe+Ni) + 16%O, (c) 84%(Fe+Ni) + 7%S + 9%H, (d) 85%(Fe+Ni) + 6%Si + 9%H. Linear combinations of models (c) and (d), and perhaps (a) and (b), are also possible. From previous experiments [70-72], it is known that model (a) also explains seismic wave velocities and the anomalous Poisson ratio of the inner core. Models (b-d) remains to be tested against these constraints, and all models need to be checked for compatibility with constraints for the liquid outer core.

**Acknowledgements**

Calculations were performed at the Supercomputing Center of Moscow State University and on Rurik supercomputer of our laboratory at MIPT. We thank the Russian Science Foundation (grant 16-13-10459) for financial support.